\documentclass[useAMS,usenatbib]{mn2e}
\usepackage{graphicx} 
\usepackage{times} %michael reckons this gives more like journal font
\def\ch2{$\chi^2$}

\def\scm  {$\hbox{{\rm cm}}^{-2}$}    %cm-2
\def \AL {$\alpha $}     %  gr. alpha
\def \HI {H{\sc \,i}}

\defcitealias{kc02}{KC03}

\def\lapp{\ifmmode\stackrel{<}{_{\sim}}\else$\stackrel{<}{_{\sim}}$\fi}
\def\gapp{\ifmmode\stackrel{>}{_{\sim}}\else$\stackrel{>}{_{\sim}}$\fi}
\def\bsp_small{\vspace{0.5cm}\small\noindent This paper has been typeset
from a \TeX/\LaTeX\ file prepared by the author.\normalsize}

\title[Detectability  of \HI\ 21-cm Absorption in DLAs]{The detectability of \HI\ 21-cm absorption in damped Lyman-{\boldmath $\alpha$}
systems}

\author[S. J. Curran and J. K. Webb]{S. J. Curran\thanks{E-mail: sjc@phys.unsw.edu.au} and J. K. Webb\\
School of Physics, University of New South Wales, Sydney NSW 2052, Australia}

\begin{document}

\date{Accepted ---. Received ---; in original form ---}

\pagerange{\pageref{firstpage}--\pageref{lastpage}} \pubyear{2006}

\maketitle

\label{firstpage}

\begin{abstract}

In this paper we investigate the possible reasons why \HI~21-cm
absorption in damped Lyman-$\alpha$ systems (DLAs) has only been
detected at low redshift: To date, no 21-cm absorption has yet been
detected at $z_{\rm abs}>2.3$ and at redshifts less than this, there
is a mix of detections and non-detections in the DLAs searched. This
has been attributed to the morphologies of the galaxies hosting the
DLAs, where at low redshift the DLAs comprise of both large and
compact galaxies, which are believed to have low and high spin
temperatures, respectively. Likewise, at high redshift the DLA
population is believed to consist exclusively of compact galaxies of
high spin temperature \citep{ck00,kc01a,kc02}. However, in a previous
paper \citep{cmp+03} we found that by not assuming or assigning an,
often uncertain, value for the coverage of the radio continuum source
by the 21-cm absorbing gas, that there is generally no difference in the spin
temperature/covering factor ratio between the 21-cm detections and
non-detections or between the low and high redshift
samples. Furthermore, only one of the 18 non-detections has a known
host morphology, thus making any link between morphology and 21-cm
detectability highly speculative.

We suggest that the lack of 21-cm absorption detections at high
redshift arises from the fact that these DLAs are at similar angular
diameter distances to the background quasars (i.e. the distance ratios
are always close to unity): Above $z_{\rm abs}\sim1.6$ the covering
factor becomes largely independent of the DLA--QSO distance, making
the high redshift absorbers much less effective at covering the
background continuum emission. At low redshift, small distance ratios
are strongly favoured by the 21-cm detections, whereas large ratios
are favoured by the non-detections. This mix of distance ratios gives
the observed mix of detections and non-detections at $z_{\rm
abs}\lapp1.6$. In addition to the predominance of large distance
ratios and non-detections at high redshift, this strongly suggests
that the observed distribution of 21-cm absorption in DLAs is
dominated by geometric effects.

\end{abstract}


\section{Introduction}\label{intro}
Redshifted absorption systems lying along the sight-lines to distant
quasars are important probes of the early to present day Universe.  In
particular, damped Lyman-$\alpha$ absorption systems (DLAs), where
$N_{\rm HI}\ge2\times10^{20}$ \scm, are useful since they account for
at least 80\% of $\Omega_{\rm neutral}$ in the Universe
\citep{phw05}. Since the Lyman-$\alpha$ transition occurs in the
ultra-violet band, direct ground based observations of neutral
hydrogen are restricted to redshifts of $z\gapp1.7$. However,
observations of the \HI ~spin-flip transitions at $\lambda_{\rm
rest}=21$ cm can probe from $z=0$, thereby providing a useful
complement to the high redshift optical data.

Provided the 21-cm and Lyman-$\alpha$ absorption arise in the same
cloud complexes, the column density $N_{\rm HI}$ [\scm] of the
absorbing gas in a homogeneous cloud is related to the velocity
integrated optical depth of the 21-cm line via
\begin{equation}
%N_{\rm HI}=-1.823\times10^{18}.T_{\rm spin}\int\!\ln\left(1-\frac {\sigma}{f.S}\right)\,dv\,,
N_{\rm HI}=1.823\times10^{18}\,T_{\rm spin}\int\!\tau\,dv\,.
\label{enew}
\end{equation}
Here $T_{\rm spin}$ [K] is the spin temperature of the gas and so in
 principle, armed with the neutral hydrogen column density from an
 observation of the Lyman-$\alpha$ line, this quantity may be derived
 from the optical depth of the 21-cm absorption.  However, the
 observed optical depth of the 21-cm line also depends upon on how
 effectively the background radio continuum is covered by the absorber
 via, $\tau\equiv-\ln\left(1-\frac{\sigma}{f\,S}\right)$,
 where $\sigma/S$ is the depth of the line relative to the
 flux density and $f$ is the covering factor of the flux by the
 absorber.

 In the optically thin regime ($\sigma/f.S\lapp0.3$), which applies to
 all but one of the known 21-cm absorbing DLAs\footnote{0235+164
 \citep{rbb+76}.}, Equation~\ref{enew} reduces to $N_{\rm
 HI}=1.823\times10^{18}\frac{T_{\rm spin}}{f}\int\!\frac
 {\sigma}{S}\,dv\,,$ thus giving a direct measure of the spin
 temperature of the gas for a known column density (from the
 Lyman-$\alpha$ line) and covering factor.  However, in the absence of
 any direct measurement of the size of the radio absorbing region,
 this latter value is often assumed or at best estimated from the size
 of the background emission region.

From the literature, 16 of the 35 DLAs searched have been found to
exhibit 21-cm absorption (see Table \ref{t1})\footnote{Since radio
frequency interference (RFI) prevented reliable observations of
0432--440 and 1228--113 \citep{cmp+03}, these are not included in the
analysis.}, all of which occur at redshifts below $z_{\rm
abs}\leq2.04$, although there are a near equal number of
non-detections also below this redshift. \citet{ck00,kc01a,kc02} therefore
advocate a scenario where low redshift DLAs have a mix of low (21-cm
detections) and high (21-cm non-detections) spin temperatures, with
the high redshift absorbers having exclusively high spin temperatures.

However, in a previous paper \citep{cmp+03}, we find evidence that the
importance of the covering factor is underestimated and the common
practice of setting $f=1$ could possibly have the effect of assigning
artificially high spin temperatures to DLAs, particularly those not
detected in 21-cm. Furthermore, we found no statistical difference in
the spin temperature/covering factor ratio between the low and high
redshift samples, although the larger absorbing galaxies (spirals)
group together at low values of ${T_{\rm spin}}/{f}$ and $z_{\rm
abs}$. Since the ratio of spin temperature/covering factor is ${T_{\rm
spin}}/{f}\propto N_{\rm HI}/\int\!\frac{\sigma}{S}\,dv$
(Equation~\ref{enew}), we have taken into account the total \HI\
column density of the absorber and integrated optical depth of the
21-cm absorption (incorporating the radio flux). This suggests that
the difference between the detections and non-detections is due
to some other effect, a possibility which we investigate in this paper.

\section{Selection effects}
\subsection{Differences between the 21-cm absorbing and non-absorbing DLAs}

We note that the cut-off of the 21-cm detections, $z_{\rm abs}=2.04$,
is close to the atmospheric cut-off of the Lyman band at $z_{\rm abs}
= 1.7$ and over the range from which the Mg{\sc \,ii} 2796/2803 \AA
~doublet may be observed by ground-based telescopes ($0.2\leq z_{\rm
abs}\leq 2.2$). Indeed only 4 of the 17 \HI\ 21-cm detections occur in
DLAs originally identified through the Lyman-\AL ~line, cf. 13 of the
18 non-detections. 
\begin{table*}
%\centering
\begin{minipage}{120mm}
\caption{DLAs and sub-DLAs searched for 21-cm absorption. As per
\citet{cmp+03}, in the top panel we list the detections and in the
bottom panel the non-detections. $z_{\rm abs}$ and $N_{\rm HI}$ are
the redshift and total neutral hydrogen column density [\scm] of the
DLA, respectively, with the optical identification (ID) given:
D--dwarf, L--LSB, S--spiral, U--unknown. The transition through which the
DLA was originally identified is given.  Finally we give
the quasar redshift and corresponding angular diameter distance
for between the quasar and absorber \citep{hog00} [throughout this paper we
use $H_{0}=75$~km~s$^{-1}$~Mpc$^{-1}$, $\Omega_{\rm matter}=0.27$ and
$\Omega_{\Lambda}=0.73$]. \label{t1} }.
\begin{tabular}{@{}l c r c c c  c r }
\hline
QSO & $z_{\rm abs}$ & $\log N_{\rm HI}$ & ID & Transition & Ref & $z_{\rm em}$ & $D_{\rm A12}$ [Mpc]\\ % $\Delta v$ [\kms]&
\hline
0235+164 &  0.52385 & 21.7 &  S & Mg{\sc \,ii} & 1 &   0.940 & 588\\ %& 64,000
0248+430 & 0.394 & -- &  U &Mg{\sc \,ii} & 2 & 1.31 & 1026 \\ %& 119,000 
0438--436$^{\dagger}$ & 2.347 & 20.8 & U &Ly-$\alpha$ & 20 &  2.852  &144 \\
0458--020 & 2.03945 & 21.7 & U & Ly-$\alpha$ & 3 & 2.286  & 98\\ %& 22,500
0738+313 & 0.2212& 20.9 & D &Mg{\sc \,ii} &  4 & 0.635   & 822\\ %&76,000
... &  0.0912 & 21.2 &  U & Ly-$\alpha$ & 5 & ...   &1119 \\ %& 99,800
0809+483$^{a}$ &  0.4369 & 20.3 &  S & 21-cm & 6 & 0.871 & 665\\ %& 70,000 
0827+243 & 0.5247 & 20.3 & S & Mg{\sc \,ii} & 7 & 0.939  & 584\\ %& 64,000
0952+179 & 0.2378 & 21.3 & L &Mg{\sc \,ii} & 2 & 1.472  & 1301\\ %&150,000 
1127--145 & 0.3127 & 21.7 & L &Mg{\sc \,ii} & 8 & 1.187 & 1093\\ %& 120,000
1157+014 & 1.94362 & 21.8&L &Ly-$\alpha$ & 9 & 1.986 & 20\\ %& 4,260
1229--021 &0.39498 & 20.8 &  S &Mg{\sc \,ii} &10 &  1.045 & 882\\ %& 95,500
1243--072 &  0.4367 & -- & S &Mg{\sc \,ii} & 11,12 & 1.286 &955 \\ %& 111,000
1328+307$^{b}$ &  0.692154 & 21.3 &  L &21-cm& 13 & 0.849  & 227\\ %& 30,000
1331+170 &1.77642 &21.2 &   U & Si{\sc \,iv} &14 & 2.084  & 146\\ %& 25,500
1629+120 & 0.5318 & 20.7 & S &Mg{\sc \,ii}& 15 &1.795  &  995\\ %& 136,000 
2351+456 &  0.779452 & --&U &21-cm  & 16 & 1.9864  & 790\\ %&121,000
\hline
0118--272 & 0.5579 & 20.3 & U &Mg{\sc \,ii}& 17 & 0.559   & 2 \\ %& 212
0201+113 & 3.386 & 21.4 & U &Ly-$\alpha$ &18 & 3.610   & 43\\ %& 16,400
0215+015 & 1.3439 &19.9 & U &Ly-$\alpha$ &19 & 1.715  & 242 \\ %& 41,000
0335--122 &3.178 &20.8 & U &Ly-$\alpha$ & 20 & 3.442  & 50 \\ %& 17,800
0336--017 & 3.0619 & 21.2 &U &Ly-$\alpha$ & 21 & 3.197  & 29 \\ %& 9650
%0438--436 & 2.347 & 20.8 & U &Ly-$\alpha$ & 20 &  2.852  &144 \\ %&  39,300
0439--433 &  0.10097 & $\sim20.0$ &  U &Mg{\sc \,ii} & 22 & 0.593  & 1048\\ %&92,700 
0454+039 & 0.8596 & 20.7 & D &Mg{\sc \,ii} &23 & 1.345  & 461 \\ %& 62,100
0528--250 & 2.811& 21.3 & U &Ly-$\alpha$ & 24 & 2.813 & 0\\ %& 157 
0537--286 & 2.974&  20.3 &U &Ly-$\alpha$ & 20 &  3.104   & 29 \\ %& 9,503
0906+430$^{c}$ & 0.63 &-- & U &Ly-$\alpha$ & 25 & 0.670  & 69 \\ %& 7,186
0957+561A & 1.391& 20.3 &U &Ly-$\alpha$ & 26 &1.413  & 17\\ %& 2,735
1225+317 &1.7941$^d$ & 19.4 &U &Ly-$\alpha$  & 27 & 2.219  & 173\\ %& 37,100
%1228--113&2.193&  20.6 &  U & Ly-$\alpha$ &20 &  3.528 &  & \\
1354--107 & 2.996& 20.8 & U &Ly-$\alpha$ &  20 & 3.006   & 2\\ %&749
1354+258  & 1.4205 & 21.5 & U &Mg{\sc \,ii} & 28 & 2.006 & 326\\ %& 60,600 
1451--375 & 0.2761 & 20.1 &U &Ly-$\alpha$   & 29 &0.314  & 100\\ %& 8,653
2128--123 & 0.4298  & 19.4 & U &Mg{\sc \,ii} &30 &  0.501  & 150\\ %& 14,191
2223--052$^{e}$ &0.4842 & 20.9 &  U &Ly-$\alpha$& 31 & 1.4040  & 941\\ %& 115,000
2342+342 & 2.9084 & 21.3 & U & Ly-$\alpha$ &18 &3.053  & 34 \\ %& 10,700
\hline
\end{tabular}
{Notes: $^{\dagger}$Just prior to this paper being accepted, a
detection of 21-cm absorption in this DLA was published
\citep{kse+06}.  This was previously flagged a non-detection by
\citet{cmp+03}, although the data and subsequent limit were poor.
$^{a}$3C196, $^{b}$3C286, $^{c}$3C216, $^{d}$21-cm absorption searched
at $z=1.795$, although the 5 MHz bandwidth used should cover
$1.781\lapp z_{\rm abs} \lapp 1.808$ \citep{bw83}, $^{e}$3C446.\\
References: $^{1}$\citet{bcs+76}, $^{2}$\citet{ss92},
$^{3}$\citet{wbt+85}, $^{4}$\citet{bktv87}, $^{5}$\citet{rt98},
$^{6}$\citet{bm83}, $^{7}$\citet{uo77}, $^{8}$\citet{bb91},
$^{9}$\citet{wmpj79}, $^{10}$\citet{kb67}, $^{11}$\citet{wpjc79},
$^{12}$\citet{wwjp83}, $^{13}$\citet{br73}, $^{14}$\citet{ysb82},
$^{15}$\citet{abe94}, $^{16}$\citet{dgh+04}, $^{17}$\citet{fal91},
$^{18}$\citet{wkb93}, $^{19}$\citet{gas82}, $^{20}$\citet{eyh+02},
$^{21}$\citet{wlfc95}, $^{22}$\citet{ptsl96}, $^{23}$\citet{bsww77},
$^{24}$\citet{smj79}, $^{25}$ \citet{wth+95}, $^{26}$\citet{ww80},
$^{27}$\citet{ulr76}, $^{28}$\citet{btt90}, $^{29}$\citet{lwt95},
$^{30}$ \cite{wwpt79}, $^{31}$\citet{lbbc93}.  }
\end{minipage}
\end{table*}
Mg{\sc \,ii} selection gives rise to a range
of absorbing galaxy types (\citealt{cks05}), and although most DLAs
discovered through the Lyman-$\alpha$ line have unidentified host
types (mainly due to the high redshift selection, e.g. Table
\ref{t1}), low redshift studies suggest that equal numbers of dwarfs
and spirals should contribute to the DLA population
\citep{rws03,zvb+05}. In Table \ref{t1} we see that the DLAs detected
in 21-cm absorption exhibit a variety of host galaxy types, although
there is the strong preference for 21-cm absorption to occur in Mg{\sc
\,ii} selected sources. However, \citet{cwm06b} find that, while the
21-cm line strength appears correlated to the rest frame equivalent
width of the Mg{\sc \,ii} line,
\begin{figure}
\vspace{6.4cm}
\includegraphics{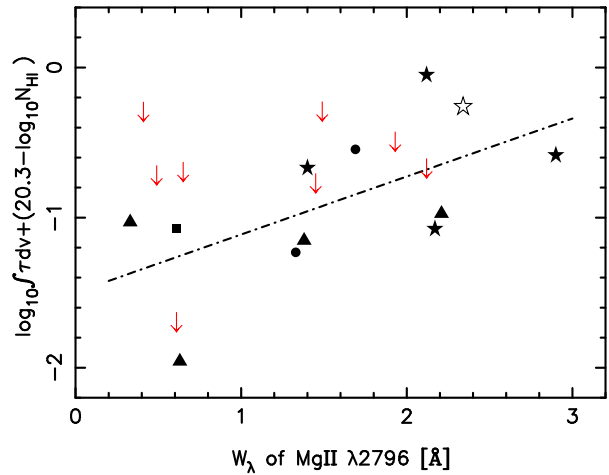}
\caption{The normalised velocity integrated optical depth
($\int\!\tau\,dv/N_{\rm HI}\propto f/T_{\rm spin}$) versus the rest
frame equivalent width of the Mg{\sc \,ii} 2796 \AA\ line. The shapes
(explained in Fig. \ref{distance}) represent the 21-cm detections
and the arrows show the upper limits for the non-detections. The line
shows the least-squares fit for the 21-cm detections. Adapted from
\protect\citet{cwm06b}.}
\label{N-W-det}
\end{figure}
21-cm absorption is perfectly detectable at low equivalent
widths. Furthermore, large equivalent widths do not necessarily ensure
a detection of 21-cm absorption (Fig. \ref{N-W-det}).

\begin{figure}
\vspace{7.85cm}
\includegraphics{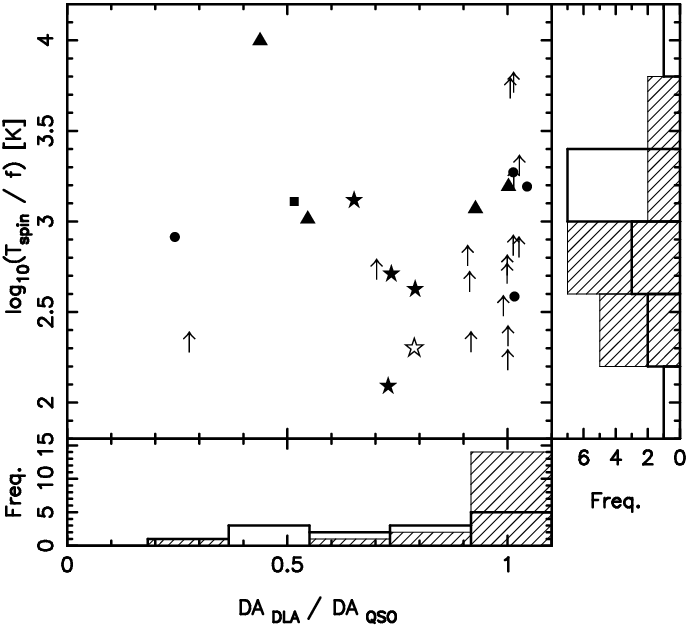}
\caption{Spin temperature/covering factor ratio versus the
absorber/quasar angular diameter distance ratio. The symbols represent
the 21-cm detections and the shapes represent the type of galaxy with
which the DLA is associated: circle--unknown type, star--spiral,
square--dwarf, triangle--LSB. The arrows show the lower limits and all
of these bar one (0454+039 at $z_{\rm abs}=0.8596$) have unknown host
identifications (Table \ref{t1}).  The unfilled star represents
0235+164 (a spiral at $z_{\rm abs}=0.524$): The 21-cm absorption in
this DLA is optically thick \protect\citep{rbb+76} and so we assume
$f=1$ which, when combined with $ N_{\rm HI}\approx5\times10^{21}$
\scm ~\protect\citep{jcb+04}, gives $T_{\rm spin}\approx200$ K.
Throughout this paper the bold histogram represents the 21-cm
detections and the hatched histogram the upper limits/non-detections.}
\label{distance}
\end{figure}

\begin{figure*}
\vspace{10.6cm}
\includegraphics{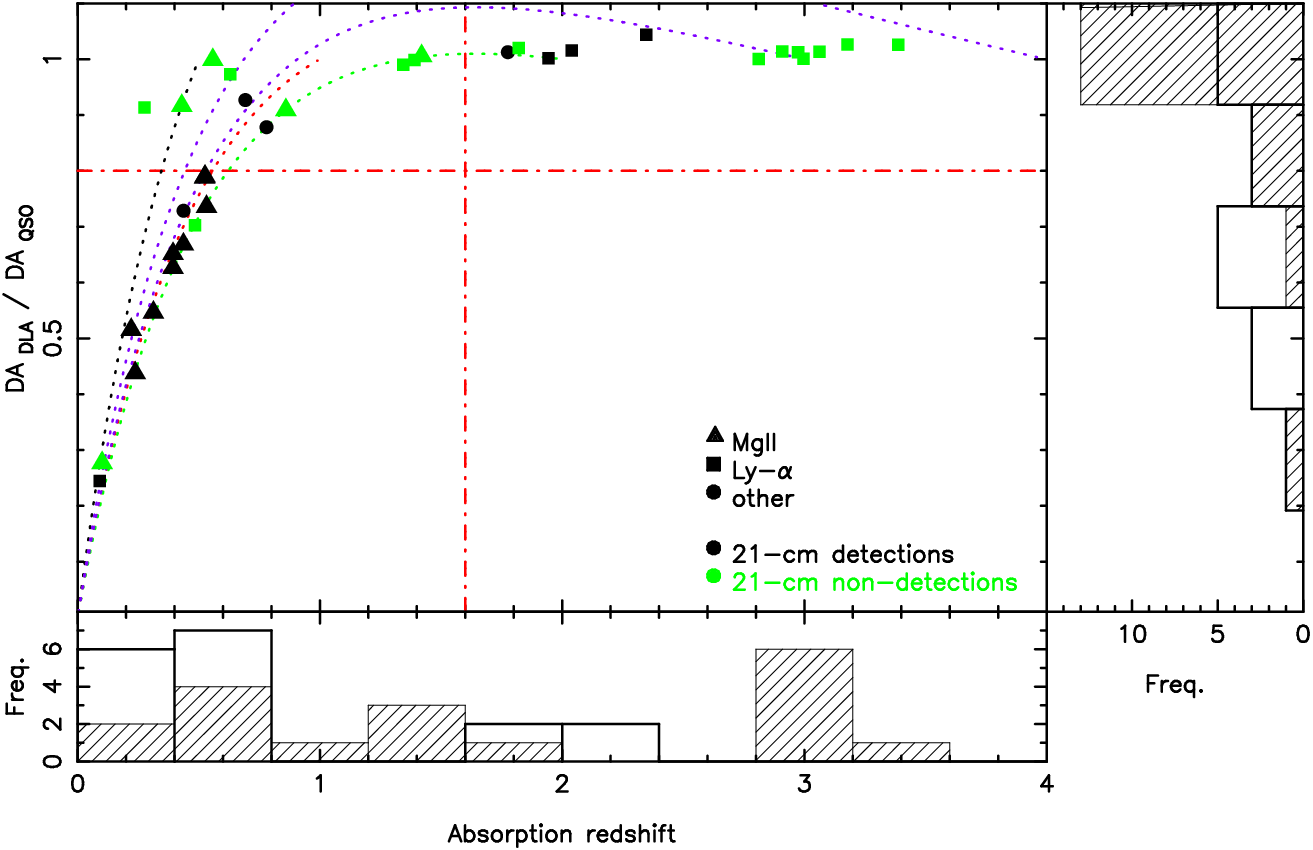}
\caption{The absorber/quasar angular diameter distance ratio versus
the absorption redshift. The black symbols represent the 21-cm
detections and the coloured symbols the non-detections, with the
shapes designating the transition through which the DLA was
discovered. Note that 0235+164 and 0827+243 are coincident at $z_{\rm
abs}\approx0.52$ and $DA_{\rm DLA}/DA_{\rm QSO}\approx0.79$. The
iso-redshift curves show how $DA_{\rm DLA}/DA_{\rm QSO}$ varies with
absorber redshift, where $DA_{\rm QSO}$ is for a given QSO redshift,
given by the terminating value of $z_{\rm abs}$. That is, we
show $DA_{\rm DLA}/DA_{\rm QSO}$ for $z_{\rm em}$ = 0.5, 1, 2, 3 and
4.}
\label{z-distance}
\end{figure*}

Since host type and Mg{\sc \,ii} equivalent width seem incidental in
determining whether 21-cm absorption is detected, we suggest that the
spin temperature/covering factor ratios in the DLAs searched for in
21-cm absorption are due to geometric effects introduced by the DLA
discovery method: In Fig. \ref{distance} we show the spin
temperature/covering factor ratio against the ratio of the angular
diameter distances\footnote{See \citet{pea99,hog00}.} to the absorber
and background continuum.
Although the sizes and morphologies of the radio sources differ
considerably (Table 2 of \citealt{cmp+03}), for given background
continuum size and 21-cm absorbing cross section, the covering factor
is obviously larger for those absorbers, at least at low redshift
(Fig. \ref{z-distance}), which are located very much closer to us than
the radio emitter. Indeed we see that the DLAs not detected in 21-cm
have significantly larger angular diameter distance ratios than the
detections, with the vast majority of these having ratios of $DA_{\rm
DLA}/DA_{\rm QSO}\gapp0.9$ (Fig. \ref{distance})\footnote{We note that
two 21-cm absorbers which are associated with LSBs, which we expect to
provide relatively small coverage, are located at a fractional
distance of $\gapp0.9$. These are 1157+014 (ratio = 1.00) and 1328+307
(ratio = 0.93). At $z_{\rm abs}=1.94$, the former is one of the
highest redshifted 21-cm absorbers known and occults a radio source
size of $<1.2$ arc-secs \citep{sfwc84} and 1328+307 occults a core
dominated source of 2.57 arc-secs \citep{vff+92}. Despite the
pitfalls in assuming given absorber and emitter sizes, for the larger
number of non-detections, the distribution does appear very skewed
towards high fractional distances.}.

In order to demonstrate the differences in fractional distances
between the 21-cm detections and non-detections, in
Fig. \ref{z-distance} we show how the distance ratio is distributed
with absorber redshift. We see that most of the 21-cm detections are
located in the bottom left quadrant, defined here by $DA_{\rm
DLA}/DA_{\rm QSO}\leq0.8$ and $z_{\rm abs}\leq1.6$. This is the
approximate redshift of the turnover in the angular diameter distance,
where objects increase their angular size with redshift. It is also
close to the redshift where the Lyman-$\alpha$ transition can be
observed by ground based telescopes, thus giving the appearance that
Lyman-$\alpha$ selected DLAs are less likely to be detected, although,
as we see from Fig. \ref{z-distance}, this is purely 
a consequence of the higher redshifts probed by this transition.

Using this partitioning, at $z_{\rm abs}<1.6$, for ratios of $DA_{\rm
DLA}/DA_{\rm QSO}<0.8$, 21-cm absorption tends to be detected (11 out
of 13 cases) and for $DA_{\rm DLA}/DA_{\rm QSO}>0.8$, 21-cm absorption
tends to be undetected (8 out of 10 cases). Within each range, if
there is an equal likelihood of obtaining either a 21-cm detection or
non-detection, the binomial probability of 11 or more out of 13
detections occurring in one bin, while 8 or more out of 12
non-detections occur in the other bin is just 0.06\%. Changing the
redshift partition to $z_{\rm abs}=1$ gives a binomial probability of
0.25\% and no redshift partition, i.e. $\geq11/13$ detections in one
bin with $\geq16/22$ non-detections in the other bin, gives
0.03\%. This leads to the hypothesis that high redshift ($z_{\rm
abs}\gapp2$) DLAs tend not to be detected in 21-cm absorption because
$DA_{\rm DLA}/DA_{\rm QSO}\approx1$ for all of these systems.

Fig. \ref{z-distance} illustrates that a redshift--distance ratio bias
arises, since at $z_{\rm abs}\gapp1.6$ the covering factor\footnote{As
defined by $f\equiv\left(\frac{r}{r_{\rm
QSO}}\right)^{2}.\left(\frac{DA_{\rm QSO}}{DA_{\rm DLA}}\right)^{2}$,
see Equation \ref{f}.} becomes effectively independent of distance and
is thereby determined by the relative extents of the absorption cross
section and continuum emission region only. At lower redshifts
(particularly $z_{\rm abs}\lapp1$) the close-to-linear decrease of
angular size with distance means that, for a given absorption cross
section, low redshift systems can much more effectively cover the
background emission. That is, above $z_{\rm abs}\approx1$, 21-cm
absorption searches are disadvantaged by the fact that in all cases
$DA_{\rm DLA}/DA_{\rm QSO}\approx1$. Furthermore, the angular
diameter--redshift relationship dictates that the bottom right-hand quadrant
of Fig. \ref{z-distance} is destined to always remain empty and the
higher the value of $z_{\rm em}$, the lower $z_{\rm abs}$ must be in
order to yield $DA_{\rm DLA}/DA_{\rm QSO}<1$.

Using the luminosity distances, a similar distribution to
Fig. \ref{z-distance} is seen, with the concentration of 21-cm
detections occurring at $D_{\rm DLA}/D_{\rm QSO}<0.5$ and the majority
of non-detections having luminosity distance ratios of $D_{\rm
DLA}/D_{\rm QSO}>0.8$. Therefore, as well as affecting the effective
coverage of the quasar's emission, the generally close DLA--QSO
proximity in the high redshift sample (Fig. \ref{distance}) could have
implications for the spin temperature of the 21-cm absorbing gas in
the DLA through increased incident 21-cm flux. We now investigate this
possibility as well as attempting to quantify the effect of the
proximity bias on the covering factors.

\subsection{Spin temperatures}
\label{st}

The high incidence of \HI\ 21-cm non-detections with large fractional
distances raises an interesting possibility: In order to explain the
decrease in the number density of Lyman-$\alpha$ absorbers as $z_{\rm
abs}\rightarrow z_{\rm em}$, against the general increase in the
number density with redshift, \citet{wcs81,bdo88} invoke a ``proximity
effect'', where at close to $z_{\rm em}$ the absorber is subject to a
high ionising flux from the quasar it occults, thus reducing the
number of Lyman-$\alpha$ clouds observed\footnote{Note that there is
also a galaxy proximity effect, where the gaseous envelopes of
galaxies close to QSOs are rarer and smaller than their QSO remote
counterparts \citep{plcw01}.}. In light of the large number of 21-cm
non-detections located relatively close to the background radio
source, an analogy of the proximity effect may be at play, where a
high 21-cm flux is maintaining a higher population in the upper
hyperfine level \citep{wb75}. This would decrease the observed \HI\
21-cm optical depth through an increase in stimulated emission
(Equation \ref{enew}).

The intrinsic luminosity of the quasar at the rest frame emission
frequency, $\nu_{\rm em}$, is $L_{\nu}=4\pi \, D_{\rm QSO}^2 \,S_{\rm
obs}/(z_{\rm em}+1)$, where $D_{\rm QSO}$ is the luminosity distance
to the quasar, $S_{\rm obs}$ is the observed flux density (given in
Table 1 of \citealt{cmp+03}) and $z_{\rm em}+1$ is the k-correction
\citep{bdo88,hog00}. Furthermore, the 1420 MHz flux density at the
absorber is $S_{\rm 1420}=L_{\nu}/4\pi \, \Delta D^2$, where $\Delta
D$ is the luminosity distance between the absorber and the
quasar. Combining this with the previous equation gives
\begin{equation}
S_{\rm 1420} = \frac{S_{\rm obs}}{(z_{\rm em}+1)}\, \frac{D_{\rm QSO}^2}{\Delta D^2} \,.
\end{equation}

Since the observed frequency is given by both $\nu_{\rm obs} = \nu_{\rm abs}
/(z_{\rm abs}+1)$ and $\nu_{\rm obs} =\nu_{\rm em}/(z_{\rm em}+1)$,
the continuum emission frequency in the rest frame of the quasar is given
by $\nu_{\rm em} = \nu_{\rm abs} (z_{\rm em}+1)/(z_{\rm abs}+1)$ 
[where $\nu_{\rm abs}=1420$ MHz in the rest frame of the absorber]. From
this, the redshift of the quasar in the rest frame of the  absorber is
given by 
\begin{equation}
\Delta z = \frac{z_{\rm em}+1}{z_{\rm abs}+1} - 1,
\end{equation}
which
we use to determine $\Delta D$. 

\begin{figure}
\vspace{7.85cm}
\includegraphics{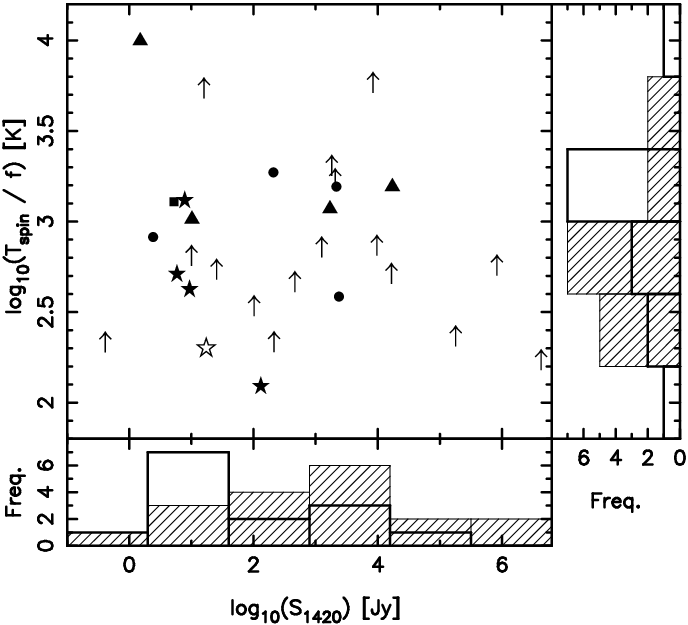}
\caption{Spin temperature/covering factor ratio versus the 21-cm flux density
at the absorber. The symbols are as per Fig. \ref{distance}.}
\label{1216}
\end{figure}

In Fig. \ref{1216} we show the observational results of the \HI\ 21-cm
searches (the spin temperature/covering factor ratio) against the
21-cm flux density calculated at a distance $\Delta D$ from the
quasar.
From this we see that below a flux density of $\sim10^{4}$~Jy at the
absorber, there is no overwhelming difference in the 21-cm detections
and non-detections, although the latter do tend to be more slightly numerous
above $\sim100$~Jy, as well as being dominant at $\gapp10^{4}$~Jy. However, the
numbers are small and it appears as though increased flux densities
due to close proximity to the background source is not the dominant
cause of the non-detection of 21-cm absorption.

\subsection{Covering factors}

Since it appears that the non-detections are not due to spin
temperatures being raised by the quasar flux, we now focus on the how the covering
factor varies with quasar proximity. As usual, we cannot separate out
the relative contributions from the spin temperature and the covering
factor, although we can define, using the small angle
approximation\footnote{Which applies to all of the background radio
sources here since $\theta_{\rm QSO}\leq64''$.}, the covering factor
as
\begin{equation}
f\equiv\frac{r^2}{DA_{\rm DLA}^2.\theta_{\rm QSO}^2},
\label{f}
\end{equation}
where $r$ is the 21-cm absorbing cross section and
$\theta_{\rm QSO}$ is the radio source size as determined from high
resolution observations (see Table 2 of \citealt{cmp+03}). It should
be borne in mind that these are usually measured at frequencies which
are several times higher than the redshifted 21-cm (1420 MHz) line.
In addition to assuming that these provide an accurate indicator of
the radio source size at the absorption frequency, Equation \ref{f}
also assumes that the emission is uniform over the extent of this
radio emission.

\begin{figure}
\vspace{7.85cm}
\includegraphics{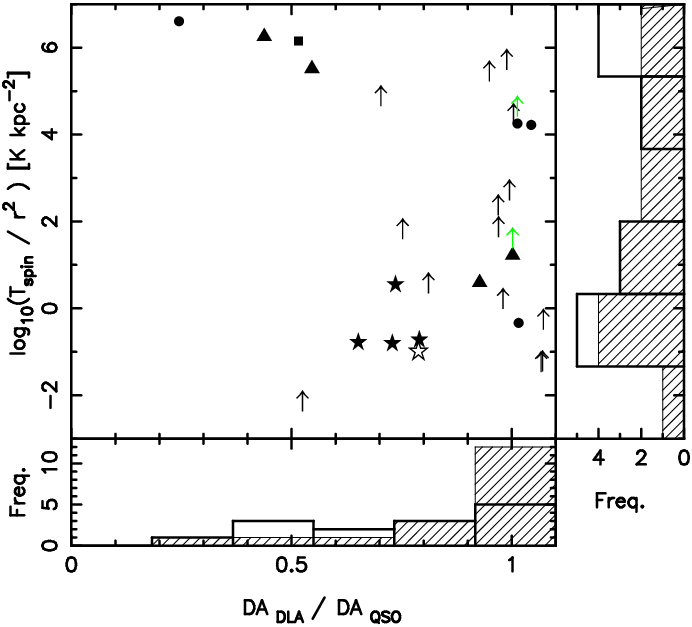}
\caption{Spin temperature/absorbing radius ratio versus the
absorber/quasar angular diameter distance ratio. The symbols are as
per Fig. \ref{distance}, with the coloured arrows designating the
lower limits due to upper limits in the radio source sizes for the
detections ($<1.2''$ and $<0.04''$, Table 2 of
\protect\citealt{cmp+03}). Note the there are also four such cases in
the 21-cm non-detections (0201+113, 0335--122, 0454+039 \& 1225+317),
but since we do not discuss these they are shown at their current
lower limits.  }
\label{Toverr-ratio}
\end{figure}

\begin{figure}
\vspace{6.4cm}
\includegraphics{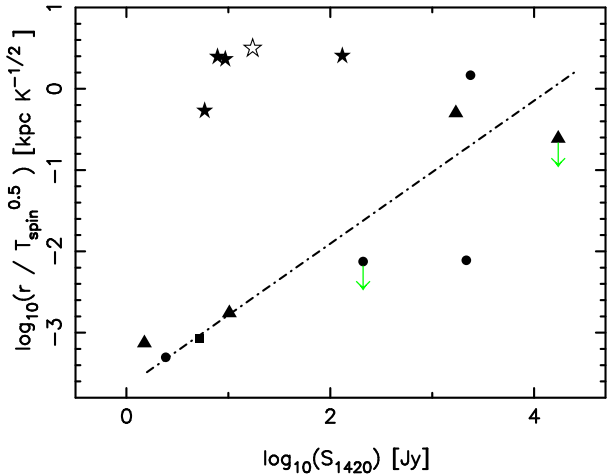}
\caption{The 21-cm absorbing cross section/spin temperature versus the
21-cm flux density at the absorber for the 21-cm detections. The
symbols are as per Fig. \ref{Toverr-ratio} and the fit is for the
seven non-spirals without upper limits and is characterised by a
gradient of 0.88 and an intercept of -3.66 (the regression coefficient
is 0.86).}
\label{r-flux}
\end{figure}

If these assumptions are reasonable, substituting Equation \ref{f}
into Equation \ref{enew} allows us to plot a covering factor ``free''
version of Fig. \ref{distance}, which we show in Fig
\ref{Toverr-ratio}. 
In the plot, like figures 4 and 5 of \citet{cmp+03}, we see a clear
distinction between the distribution of the spirals and the more
compact galaxies (more evident in Fig. \ref{r-flux}).  This may
indicate that, at least at low redshift (where $D_{\rm DLA}/D_{\rm
QSO}<0.8$), each group has similar ratios ($T_{\rm spin}/r^2\sim10^6$
K kpc$^{-2}$ for the compact galaxies and $\sim0.1$ K kpc$^{-2}$ for
the spirals), with the absorbing cloud size making a large
contribution in the very different values between these two groups: A
span of $\approx7$ orders of magnitude seems unlikely through spin
temperature alone, although for a given temperature, a span of only
$\sim3$ dex is required in the radius of the absorbing
region. Furthermore, figure 4 of \citet{cmp+03} shows that the radio
sources of $\lapp0.1''$ tend to be adequately covered by the compact
galaxies, whereas for the larger radio sources spirals are required.

Bearing in mind that two of the $T_{\rm spin}/r^2$ values at $DA_{\rm
DLA}/DA_{\rm QSO}\approx1$ ($D_{\rm DLA}/D_{\rm QSO}\gapp0.8$) are
lower limits, in Fig. \ref{Toverr-ratio} there may be a trend for
$T_{\rm spin}/r$ to decrease as the DLA--QSO distance (angular
diameter \& luminosity) closes, for 21-cm absorption detected in
non-spirals.  Presuming that the spin temperature does not decrease
with proximity to the quasar (contrary to what we would
expect)\footnote{This suggests that the spin temperature also
decreases with redshift, contrary to the results of \citet{kc02}.},
this may suggest a selection effect where only large 21-cm
absorbers are detected close to the background continuum, implying
that self shielding against high fluxes are important, where the
effectiveness of this scales with cloud size (Fig. \ref{r-flux}).
This, however, relies upon the aforementioned assumptions regarding
the radio sources and many more detections would be required to
adequately test this hypothesis. We note with interest, that the 21-cm
detection located furthest to the bottom right in
Fig.~\ref{Toverr-ratio} (absorber ID unknown at $DA_{\rm DLA}/DA_{\rm
QSO}=1.02$), which has the very low value of $T_{\rm spin}/r^2=0.46$ K
kpc$^{-2}$ ($r/\sqrt{T_{\rm spin}}=1.5$ kpc K$^{-1/2}$,
Fig. \ref{r-flux}), is due to 0458--020, where a large absorbing cross
section of $r>10$ kpc is deduced from VLA and VLBI observations
(\citealt{bwl+89}).

\section{Summary}

Regarding the detectability of 21-cm absorption in DLAs, we have found:
\begin{itemize}
  \item In general, the
non-detections of 21-cm absorption in DLAs have been searched as
deeply as the detections, meaning that the ratio of spin
temperature/covering factor does not differ significantly between the
two samples.

\item There is an apparent bias for 21-cm absorption to be detected in
DLAs originally discovered through the Mg{\sc \,ii} doublet rather
than the Lyman-\AL\ line. This, however, is superficial and merely
reflects the true bias introduced by the redshift distribution of the
DLAs:
\end{itemize}
At $z_{\rm abs}\gapp1.6$, the absorbers are effectively at the same
(angular) distances as the background quasars. After ruling out the
possibility that the closer proximity of the undetected DLAs to the
background quasars significantly raises the spin temperatures above
those of the detections, we believe that the non-detections are due to
low covering factors, the result of the flattening of the angular
diameter distance at $z\gapp1$.

Since DLAs are not detected in 21-cm absorption at $z_{\rm
	abs}\geq2.04$\footnote{The highest redshift of a confirmed
	detection is now $z_{\rm abs}=2.347$ (Table \ref{t1}).},
	\citet{ck00,kc01a,kc02} suggested that the non-detections are due to
	the high redshift ($z_{\rm abs}\gapp2$) DLAs having
	exclusively high spin temperatures and the presence of both
	21-cm detections and non-detections at low redshift is
	attributed to a mix of spin temperatures. However, the
	distribution of 21-cm detections and non-detections closely
	follows that of low and high angular diameter distance ratios,
	respectively, so that high redshift DLAs have exclusively high
	ratios, whereas $z_{\rm abs}\lapp2$ systems exhibit a mix of
	ratios. This geometric effect means that absorbers at high
	redshift will always cover the background quasar much less
	effectively than at low redshift and the degeneracy between
	spin temperature and covering factor may only ever be resolved
	by targetted searches for 21-cm absorption in high redshift
	DLAs towards very compact radio sources.

\section*{Acknowledgments}

We would like to thank Matthew Whiting and Michael Murphy for their
advice as well as the referee, Emma Ryan-Weber, for her prompt,
detailed and very helpful review in addition to her subsequent
feedback. This research has made use of the NASA/IPAC Extragalactic
Database (NED) which is operated by the Jet Propulsion Laboratory,
California Institute of Technology, under contract with the National
Aeronautics and Space Administration.  This research has also made use
of NASA's Astrophysics Data System Bibliographic Services.

%\bibliographystyle{mn2e}
%\bibliographystyle{apj} %same problem as before (m.tex)
%\bibliography{../../bib/aa,../../bib/ref} %set at Nobeyama because not working
%\bibliography{aa,ref}
%\expandafter\ifx\csname natexlab\endcsname\relax\def\natexlab#1{#1}\fi

\label{lastpage}
\end{document}